\documentclass[twocolumn,showpacs,showkeys,preprintnumbers,amsmath,amssymb,a4paper]{neel}
\usepackage{graphicx}
\usepackage{dcolumn} 
\usepackage{multirow}

\begin{document}

\DocReference{Document de travail}
\PageWeb{http://crtbt.grenoble.cnrs.fr/helio/}
\DateDerniereCompilation{Derni\`ere compilation : \today}
\PSLogo{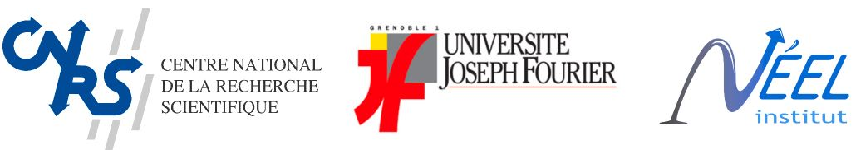}
\LogoHeight{1.7cm}

\title{Evidence of a boundary layer instability at very high Rayleigh number}

\author{F. Gauthier and P.-E. Roche}

\affiliation{Institut NEEL, CNRS/UJF\\
BP 166, F-38042 Grenoble Cedex 9, France}

\date{\today}

\begin{abstract}
In 1997, a Rayleigh-B\'enard experiment evidenced a significant increase of the heat transport efficiency for Rayleigh numbers larger than $Ra \sim  10^{12}$ and interpreted this observation as the signature of the Kraichnan's ``Ultime Regime'' of convection. According to Kraichnan's 1962 prediction, the flow boundary layers above the cold and hot plates -in which most of the fluid temperature drop is localized- become unstable for large enough $Ra$ and this instability boosts the heat transport compared to the other turbulent regimes. Using the same convection cell as in the 1997 experiment, we show that the reported heat transport increase is accompanied with enhanced temperature fluctuations of the bottom plate, which was  heated at constant power levels.
 Indeed, for $Ra < 10^{12}$, the bottom plate fluctuations can simply be accounted from those in the bulk of the flow. In particular, they share the same spectral density at low frequencies, as if the bottom plate was following the slow temperature fluctuations of the bulk, modulo a constant temperature drop across the bottom boundary layer. Conversely, to account for the plate's temperature fluctuations at higher $Ra$, we no-longuer can ignore the fluctuations of the temperature drop across the boundary layer. The negative skewness of fluctuations at high $Ra$ supports the picture of a boundary layer instability.
These observations provide new evidence that the transition reported in 1997 corresponds to the triggering of the Ultimate Regime of convection.
\end{abstract}



\maketitle


\section{Introduction}

Natural convection forces turbulence in the atmosphere, in the oceans as well as in numerous geophysical, industrial and astrophysical flows. Understanding the convection regimes settling on such large scales is a major scientific challenge, directly impacting our ability to assess heat and mass transfer in the environment, for example. One of the simplest laboratory paradigm to explore the basic mechanisms of turbulent convection is the Rayleigh-B\'enard cell. It consists in an horizontal layer of fluid between two isothermal plates imposing a temperature difference $\Delta$ destabilizing the fluid. Dimensional analysis shows than the main control parameter in such cells is the Rayleigh number ($Ra$) which is proportional to $\Delta$ and to the cube of the fluid layer height  (the complete definition of $Ra$ is given later). Unfortunately, due to this geometrical dependence, the reachable Rayleigh numbers in laboratory experiments are decades smaller than those estimated for environmental flows (typically  $Ra \sim 10^{20} $ in the atmosphere for example). Thus, our confidence in extrapolating laboratory findings to such flows depends on the understanding of very high $Ra$ experiments.

Since 1997, a few Rayleigh-B\'enard experiments \cite{chavanne1997,RochePRE2001,Niemela2003,RochePoF2005} reported a significant enhancement of heat transport efficiency across the cell above $Ra \sim 10^{12}$ within Boussinesq conditions. These observations were interpreted as the signature of the ``Ultimate Regime'' of convection, predicted in 1962 by R. Kraichnan \cite{Kraichnan1962}. Qualitatively, this regime is characterized by the turbulent state of the two boundary layers laying close to each plate but the precise conditions for triggering this boundary layer instability are difficult to predict \cite{Kraichnan1962}. The interpretation presented in 1997 \cite{chavanne1997} was motivated by the observation of an unprecedented heat transfer scaling versus $Ra$, compatible with the onset of Kraichnan's regime and significantly more efficient than any other observed or predicted scaling. In another paper, the authors of \cite{chavanne1997} reported another signature of the transition on the local temperature statistics at mid-height within the cell  \cite{Chavanne2001}. In 2001,  the observation of the asymptotic heat transport scaling of the Ultimate Regime was reported in a similar cell with a calibrated corrugated surface \cite{RochePRE2001}. For usual velocity turbulent boundary layers, such surfaces are well known to set the viscous sub-layer height constant and thus to reveal the asymptotic transport scaling \cite{Schlichting}. This experiment therefore supports the 1997 interpretation of the occurrence of the Ultimate Regime of convection.

But a more direct evidence of a laminar-to-turbulent transition of the boundary layers near  $Ra =10^{12}$ is still missing, which motivated the present work. Direct local measurements within the boundary layers are difficult for practical reasons : the thermal boundary layer is very thin (typically smaller than 100 $\mu m$ in the experiments discussed above) and its temperature profile is Ra dependent. To bypass the difficulties of local invasive measurements within the boundary layer, we use the low frequency fluctuations of the bottom plate to probe the stability of the bottom boundary layer. More precisely, we apply a constant heating power $J$ on the bottom plate and monitor its temperature fluctuations of order one percent of $\Delta$. We focus on times scales longer than the turn-over time of the mean flow circulation, for which the lack of spatial resolution of the ``plate-probe'' is not a limitation. We find that above $Ra \sim 10^{12}$, such fluctuations become significantly larger than those in the bulk of the flow, while both have the same intensity for lower $Ra$. Besides, the probability density function of the bottom plate temperature  becomes increasingly skewed towards negative values above $Ra \sim 10^{12}$, corresponding to more and more frequent intense cooling events. We show that both observations are straightforwardly interpretable by the occurrence of an instability in the boundary layer for $Ra \sim 10^{12}$, which is consistent with the scenario of a transition to the Ultimate Regime of convection.

For reference, we mention that three Rayleigh-B\'enard experiments explored $Ra$ larger than $10^{13}$ and didn't find any significant enhancement of the heat transfer \cite{WuTHESE,Ashkenazi1999,niemela2000}. In \cite{WuTHESE}, a transition of the local temperature statistics have been reported for $Ra \simeq 10^{11}$ and $10^{13}$ \cite{Procaccia1991}. In \cite{Niemela:2006}, the authors report an enhancement of heat transport at high $Ra$ but, according to the authors themselves, the data are not fulfilling the Boussinesq conditions  and therefore are difficult to compare to others. For references on 
numerical simulations, see \cite{Kenjeres2002,Amati2005}. The reason for such a scatter between experiments at very high $Ra$ is still not understood and this issue is not addressed in this paper. We just note that the laminar - turbulent transition of  velocity-driven boundary layers  is known to be highly sensitive to experimental conditions \cite{LandauMecaFlu_English} and the same could remain true for ``temperature-driven'' boundary layers.

For convenience, we recall here the definitions of the main dimensionless parameters. The Rayleigh number $Ra=(g h^3 \alpha \Delta)/(\nu \kappa)$ and the Prandtl number $Pr = \nu / \kappa$ characterize the buoyancy force  and the molecular transport properties where $g$ is the gravity acceleration, $\Delta$ is temperature difference across the cell of height  $h$ and $\alpha$, $\nu$, $\kappa$ are respectively the isobaric thermal expansion coefficient, the kinematic viscosity and the thermal diffusivity.  The Nusselt number $Nu = J h / (S k \Delta)$ characterizes the heat transport efficiency across the cell, where $J$ is the mean heat flux (in practice, the applied Joule heating), $k$ the thermal conductivity of the fluid and $S$ the surface of the plate-fluid interface.

\section{Experimental setup}

The convection apparatus used in the present work has been described in \cite{Chavanne2001}. The fluid is confined in a cylindrical volume of height $20\ cm$  and diameter $10\ cm$ with top and bottom oxygen free high conductivity copper (OFHC) plates of thermal conductivity $1090\ W.m^{-1}.K^{-1}$ at $4.2\ K$ \cite{gauthierETC11_2007}. The seamless stainless steel side wall is $500\ \mu m$ thick and its measured thermal conductance is $327\ \mu W . K^{-1}$ at $4.7 K$. Its parasitic contribution on $Nu$ has been corrected using the correction formula proposed in \cite{RocheEPJB2001}. The cell is hanging in a cryogenic vacuum : the top plate is cooled by an helium bath at $4.2\ K$ through a calibrated thermal resistance ($3\ W/K$ at $6\ K$). Its temperature is regulated by a PID controller 
 and we apply a constant and distributed Joule heating 
on the bottom plate. The heat leaks from the cell in such set-up (about $200\ nW$ at $4.7\ K$) are at least 4 decades smaller than this heating and are essentially due to the radiative transfer as shown in \cite{chavanne1996}. The temperature difference $\Delta$ between the plates is measured with $0.1\ mK$ accuracy thanks to a thermocouple ; for comparison the smallest $\Delta$ used in this work is $18\ mK$. The cell is filled with $^{4}He$ which properties are calculated based on Arp \& McCarty works \cite{McCarty1990} and later improvements \cite{RocheJLTP2004}. The cell has been operated for $3$ mean density and temperature conditions : $19.15\ kg/m^3$ at $6.0\ K$ ($Pr\simeq 1.0$),  $19.15\ kg/m^3$ at $4.7\ K$ ($Pr\simeq 1.3$) and $70.31\ kg/m^3$ at $6.0\ K$ ($Pr\simeq 2.8$). These values are a compromise to obtain the best accuracy of helium properties, a 3 decades range of $Ra$ around $Ra=10^{12}$ and a limited range of $Pr$.

We now show that the temperature of the bottom plate $T_{plate}$ is a well defined and measurable global quantity in such a set-up, accounting for the homogeneous temperature fluctuations of the whole plate. Two germanium thermistances are inserted into thermometers holders screwed on this copper plate \footnote{Two different types of thermistances are used on each plate : LakeShore $GR-200B-2500$ and Cryocal $CR1500-PB$. Special attention is dedicated to the thermalization using standard cryogenics techniques (Apiezon contact grease, golden copper surfaces, etc...).}. The thermal response time of each thermistance is measured with the $3\omega$ technique (for example see \cite{Lu2001}). We find cut-off frequencies at $-3\ dB$ between $16\ Hz$ and $32\ Hz$  which is consistent with the thermistances' specifications. Thus thermometry has enough dynamics to reflect the plate's temperature fluctuations up to a few $Hz$, which is above frequencies of interest.
The main intrinsic response time of the plate is $\tau_{p} = d_{p}^{2}/\kappa_{p}$ where $d_{p}=10\ cm$ is the plate's diameter and $\kappa_{p}$ is the thermal diffusivity of copper. At $4.7\ K$, $ \tau_{p}^{-1} $ is about $120\ Hz$ which is also larger than the frequencies of interest. Although the plate's diffusive response time is small, the non-homogeneous non-stationary heat fluxes at the fluid/plate interface can still generate temperature gradients  in the plate (for example, see \cite{Chilla2004}). We checked that this effect was negligible for the frequency range of interest : the coherence function between one thermometer located on the cell axis and the other one located on the plate edge is larger than $97\%$ (or $99\%$ if we discard the 3 lowest $\Delta$). $T_{plate}$ is therefore a measurable and  well defined quantity in this set-up, at least up to a few Hz.

\begin{figure}[!ht]
\includegraphics[width=\linewidth]{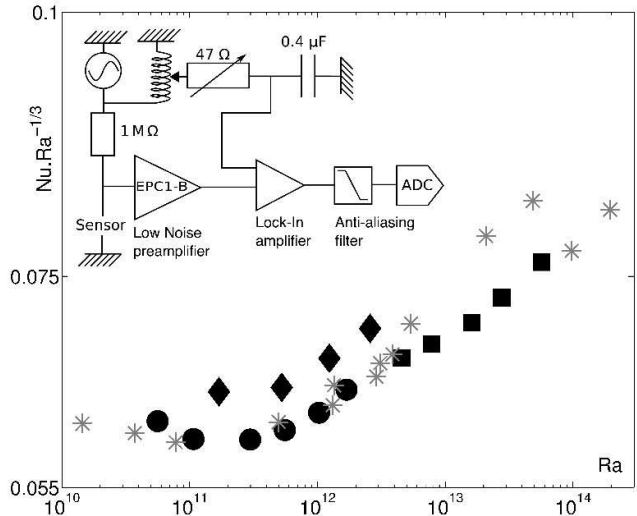}
\caption{Compensated Nu versus Ra. Solid symbols : present work. $\bullet$ : Pr=1.0, $\blacklozenge$ : Pr=1.3, $\blacksquare$ : Pr=2.8. $\ast$ : data from \cite{Chavanne2001}. Insert : measuring circuit of each thermistance sensor.}
\label{germanium}   
\end{figure}

Each thermistance is polarized with an AC current of a few $\mu A$ (to prevent significant overheating) and the resulting voltage is demodulated with a lock-in amplifier of $20\ ms$ time constant, connected in a bridge configuration (insert of Fig.\ref{germanium}). The output signal is anti-alias filtered and recorded continuously during 5 or 10 h, corresponding typically to a few thousands of turn-over times.

\section{Results}

As a preliminary test, we checked that the heat transport law $Nu(Ra)$ across the cell is consistent with the previous measurements with the same apparatus \cite{Chavanne2001}, in particular regarding the occurrence of a heat transfert enhancement slightly below $Ra = 10^{12}$. Figure \ref{germanium} illustrates this agreement by representing the compensated heat transfer $Nu Ra^{-1/3}$versus $Ra$ for both experiments (data from \cite{Chavanne2001} are restricted to those for which temperature time series in the bulk are available).

\begin{figure}[!ht]
\includegraphics[width=\linewidth]{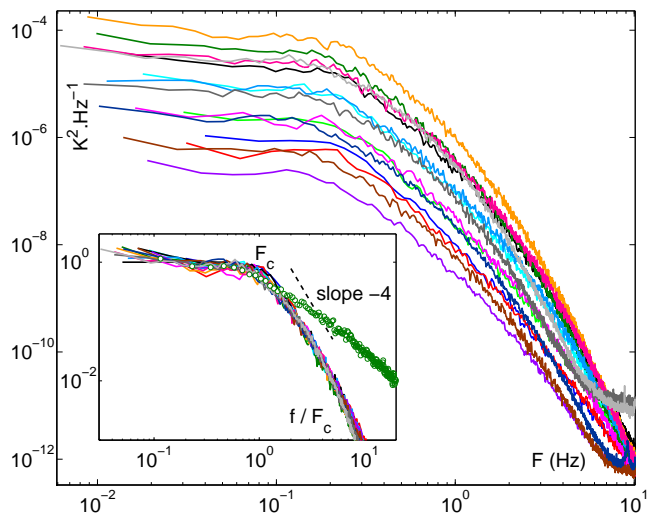}
\caption{Power spectra of the bottom plate temperature $T_{plate}$. Insert : same spectra normalized by the adjusted plateau value $P$ (y axis) and by $F_{c}$ (x axis).  $\circ$  is a power spectrum of $T_{bulk}$ for $Ra = 1.35 \times 10^{12}$ recorded in a previous run. }
\label{spectrebrut}   
\end{figure}

Figure \ref{spectrebrut} shows the power spectrum density of the bottom plate temperature $T_{plate}$ for various $Ra$.
At low frequencies,  spectra saturate on what we shall refer as ``plateaus''. These plateaus  will be characterized by their spectral densities $P$.
The cut-offs at high frequency are steeper than -3 dB/dec and they can be characterized by the frequencies $F_{20dB}$ at -20 dB below the plateaus, or more conveniently by $F_c=0.2 \times F_{20dB} $ which roughly corresponds to the cross-over between the plateau and the cut-off regions.
In pratice, $P$ has been determined as the value ensuring the best merging of all spectra, when the spectral density is divided by $P$ and the frequency by $F_{c}$ (solid lines Fig. \ref{spectrebrut} insert).
Dimensionless cut-off frequencies $F_c^{\star}$ and plateaus $P^{\star}$ are defined by :
\begin{equation}
F_c^{\star} = F_c . h^2 / \nu \,\,\,\,and \,\,\,\, P^\star = P . \nu /(h^2 \Delta ^2)
\end{equation}

\begin{figure}[h]                                                                                                                                                                  
\includegraphics[width=\linewidth]{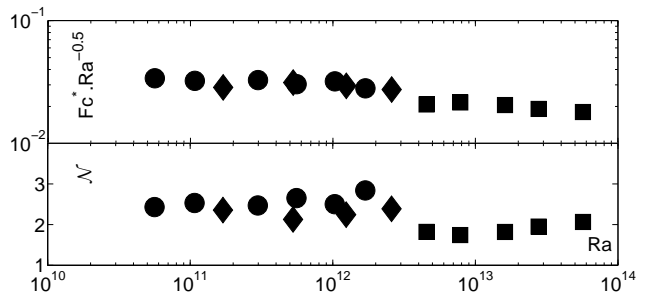}
\caption{Upper plot : compensated cut-off frequency $F_{c}^\star$ versus the Rayleigh number. Lower plot : estimated number of turn-overs corresponding to $F_{c}$}
\label{fc}   
\end{figure}

$F_c^{\star}$ can be seen as the Reynolds number associated with the distance $h$ and the velocity $h F_{c}$. Figure \ref{fc} presents this quantity compensated by $Ra^{0.5}$  to illustrate that the scaling of $F_c^{\star}$ is close to a power law $Ra^{1/2}$ for a given $Pr$. Similar scaling have already been reported for other Reynolds numbers \cite{Chavanne2001,Niemela2001} measured in the same window of $Ra$. In particular, the Reynolds number $Re$ associated with a mean velocity $V$ of the large scale circulation at mid-height (at $2.5\ cm$ from the axis) has been measured and fitted in the present cell as \cite{Chavanne2001} :
\begin{equation}
Re = h V / \nu \simeq  0.206 \times Ra^{0.49} Pr^{-0.70}
\end{equation}

The lower plot of Fig. \ref{fc} gives an estimate of the number $\cal{N}$ of turn-overs of the mean flow circulation during a time period $1/F_{c}$. Assuming that the circulation path is $2h$ long and using the previous mean velocity $V$ fit, we get $N \simeq V /2 h F_{c} = Re / 2 F_{c}^{\star}$. The insert shows that $\cal{N}$ is close to 2, indicating that the cross-over between the plateaus and the cut-off region of the $T_{plate}$ spectra is of the order of one turn-over frequency.

\begin{figure}[h]
\includegraphics[width=\linewidth]{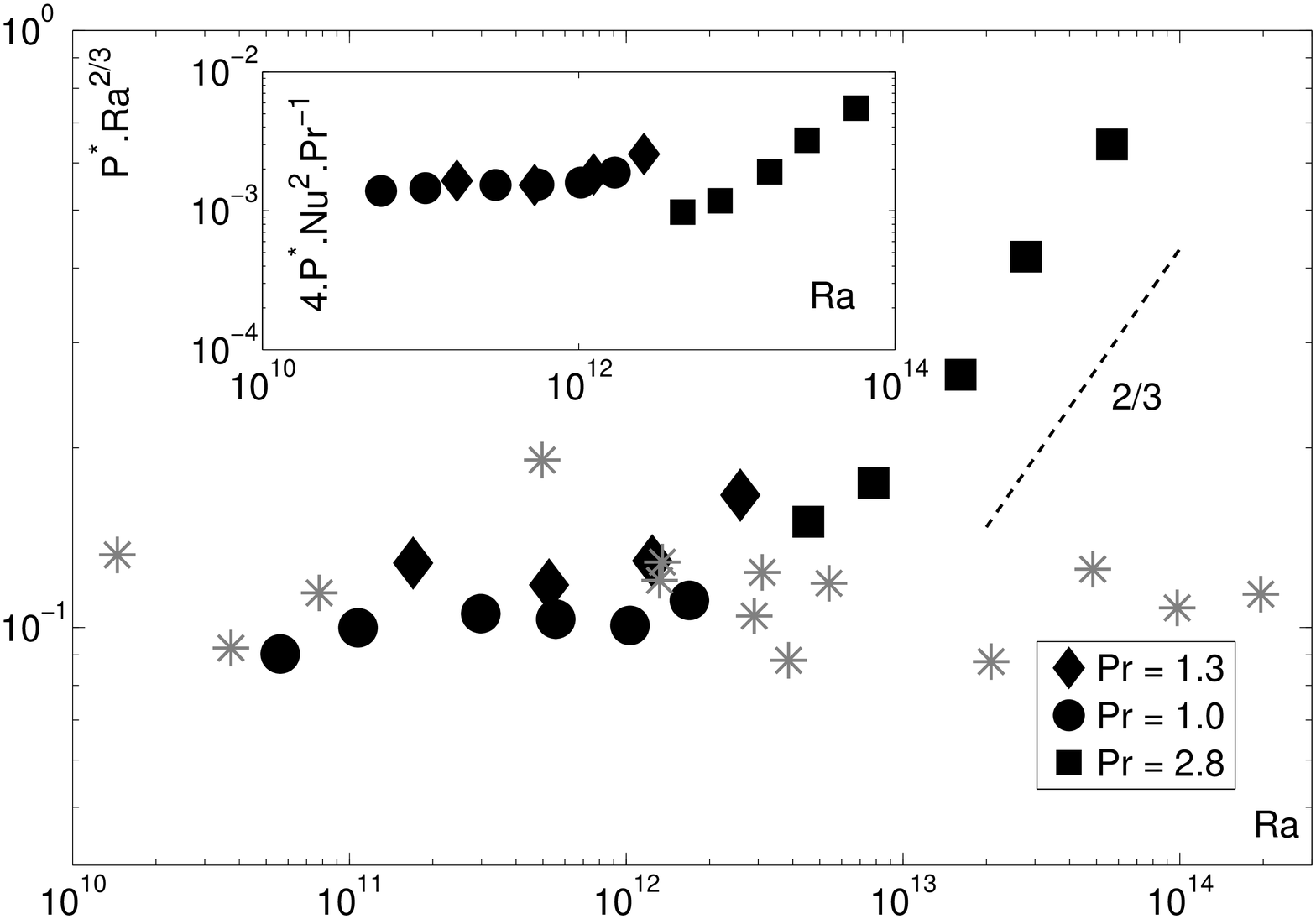}
\caption{Compensated plateau level $P^{*}$ versus Rayleigh number (same symbols as on Fig.1). Insert : alternative normalisation of $P$ (see text).}
\label{plateau}   
\end{figure}

The compensated plot of the spectral densities $P^{\star}$ on Fig. \ref{plateau} is the first important result of this paper. For $Ra<10^{12}$, the spectral densities dependence $P^{\star} (Ra)$ is compatible with a $Ra^{-2/3}$ scaling while it has no obvious $Ra$ dependence for $Ra > 2 \times 10^{12}$, as illustrated by the $2/3$ slope (dash line).
For comparison, we also plotted the corresponding spectral density calculated from temperature time series  recorded at mid-height within the cell in the previous experiment \cite{Chavanne2001}. Similar spectral plateaus can also be evidenced in these times series below the frequency corresponding to a turn-over time typically. Within experimental uncertainty, the spectral densities plateaus in the bulk overlap with the plate's ones below $Ra=10^{12}$, but they significantly diffe r above $Ra=10^{13}$. The insert shows $4 P^\star Nu^2 Pr^{-1}$ versus $Ra$. This quantity can be seen as $P$ made dimensionless with $\Delta$ and the molecular thermal diffusion time across a boundary layer of heigh $h/(2 Nu)$. Within uncertainty, such a compensation of $P^\star$ also cancels the $Ra$ -and maybe $Pr$- dependences in the low $Ra$ region.

\begin{figure}[h]
\includegraphics[width=\linewidth]{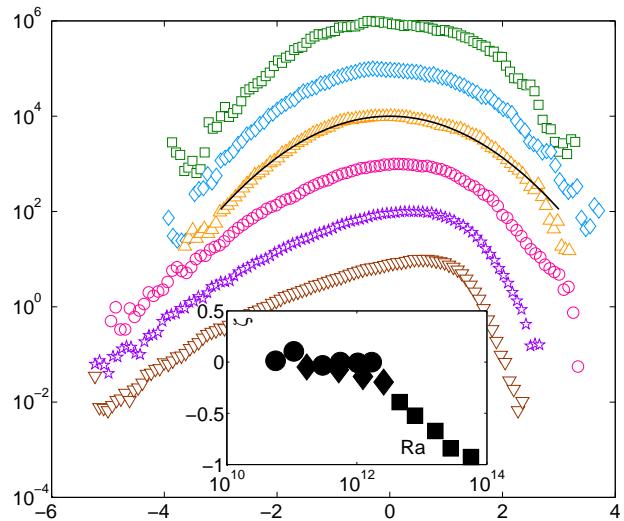}
\caption{Probability density functions of $T_{plate}-\overline{T_{plate}}$ normalised by its standard deviation. A vertical offset of one decade is introduced between each datasets. From top to bottom : ($\square$) $Ra = 5.6 \times 10^{10}$,($\diamond$) $Ra = 5.5 \times 10^{11}$,($\bigtriangleup$) $Ra =1.7 \times 10^{12}$,($\circ$) $Ra =4.5 \times 10^{12}$,($\star$) $Ra =1.6 \times 10^{13}$,($\bigtriangledown$) $Ra =5.7 \times 10^{13}$. The insert shows the skewness versus Ra (same symbols as on Fig.1)}
\label{hist}   
\end{figure}

Figure \ref{hist} shows a representative sample of the probability density functions (pdf) of $ T_{plate} -\overline{T_{plate}} $ normalized by its standard deviation, where overline stands for time averaging. The Rayleigh numbers range from $5.6 \times 10^{10}$ (top) up to $5.7 \times 10^{13}$ (bottom) where vertical offsets are introduced for clarity. The first three pdf are close to Gaussian (the solid line corresponds to a gaussian distribution), while the others become increasingly skewed as $Ra$ increases : intense coolings of the plate occur
more and more frequently than the corresponding overheating events. The insert presents the skewness
\begin{equation}
\zeta = \frac{ \overline{( T_{plate} - \overline{T_{plate}} )^3} }{\overline{( T_{plate} -\overline{T_{plate}} )^2}^{3/2}}
\end{equation}

 \noindent versus $Ra$. It
illustrates that the absolute value of the skewness is no longuer significantly smaller than unity above $Ra \sim 10^{12}$.

\section{Discussion}

We assume that, on time scales longer than a few turn-over times, the temperature fluctuations locally recorded in the bulk of the flow reflect temperature variations spanning over the whole bulk of the flow. Accordingly, we interpret  the spectral density plateaus of $T_{bulk}$ below $F_{c}$ as a global characteristics of the bulk temperature, rather than local one. Since we operate the bottom plate at constant heating power, its temperature $T_{plate}$ is referenced by the fluid's bulk temperature and the temperature drop across the bottom boundary layer. At frequencies smaller than $F_{c}$, we can define a time-dependent temperature drop $\delta T_{bl}$ across the bottom boundary layer, which mean value is $\Delta /2$ and which  fulfills (by definition)\,:
\begin{equation}
T_{plate}= T_{bulk} + \delta T_{bl}
\end{equation}

Below $Ra \simeq 10^{12}$, we conjecture that the bottom plate temperature fluctuations follow the ones of the bulk with negligible noise introduced by the temperature drop $\delta T_{bl}$ across the bottom boundary layer, that is :
\begin{equation}
T_{plate} \simeq T_{bulk} + \Delta T /2
\end{equation}

This is supported by the overlapping spectral densities plateaus of $T_{bulk}$ and $T_{plate}$ reported on Fig. \ref{plateau}. 
The open circle of the insert of Fig. \ref{spectrebrut} presents a typical spectrum of the temperature $T_{bulk}$ for $Ra=1.35 \times 10^{12}$ and $Pr=1.3$ recorded during the previous experiment with the same cell. The Y axis rescaling has been done with the same procedure as previously. The frequency axis is rescaled  by $F_c=0.068\ Hz$, which was obtained from an interpolation between the solid symbols of Fig. \ref{fc}. The cut-off of the $T_{bulk}$ spectra for $F>F_{c}$  is less steep than for $T_{plate}$ spectra. This is consistent with a spatial filtering by the plate of the fluid's temperature inhomogeneities extending over less than typically one cell diameter. Estimations of the thermal characteristics of the fluid-plate system also corroborate our conjecture. Indeed, the expected time response of the plate to a slow variation of $T_{bulk}$ can be estimated as $RC$, where $C \simeq 1 J/K$ is the (measured) plate's heat capacity and $R$ is the bottom boundary layer dynamical thermal resistance $R$   :
\begin{equation}
R= \frac{\partial \Delta /2}{\partial J} \simeq \frac{1}{1.3} \frac{\Delta /2}{J}
\label{resdyn}
\end{equation}

We find that $RC$ is always more than one decade smaller than $1/F_c$. Since the heat capacity of the bottom plate is significantly smaller than the heat capacity of the fluid, the plate represents a negligible thermal load for the slow temperature variations of the bulk fluid. Thus the plate can indeed follow quasistatically the temperature variation of the bulk, modulo the offset introduced by the temperature drop across the boundary layer.

Above  $Ra \simeq 2 \times 10^{12}$,  the spectral density plateaus $P^\star$ of $T_{plate}$ get increasingly larger than their counterpart in the bulk (Fig. \ref{plateau}).  We interpret this strong enhancement of fluctuations in the plate as the contribution of the temperature drop $\delta T_{bl}$ across the boundary layer. At high $Ra$, the frequent intense cooling events evidenced with the probability distributions of Fig. \ref{hist} are interpreted considering that the  thermal contact between the plate and the bulk is  greatly enhanced during short periods of time, corresponding to brief thinning of the boundary layer. This strongly suggests that the boundary layer is unstable.

We now discuss and discard the possibility that this boundary layer instability was already present at lower $Ra$ but that the corresponding temperature fluctuations of $\delta T_{bl}$ were simply masked by the fluctuations of the bulk $T_{bulk}$. Within this hypothesis of this ``masking-effect'', Fig. \ref{plateau} could be interpreted without resorting to a flow transition occurring near $Ra \sim 10^{12}$.

A first argument is that both enhancements of Nu and $T_{plate}$ fluctuations nearly occur for the same $Ra$. If we accept that this coincidence is not fortuitous, both events should be accounted by a single mechanism which cannot be a simple masking-effect hypothesis.


As a second argument, the small -if any- dependence of $P^\star (Ra)$ for $Ra>10^{12}$, seems in contradiction with our knowledge of boundary layers physics in the hard turbulence regime, which preceeds the Ultimate one. This contradiction suggests that the observed high $Ra$ regime is not the hard turbulence one.
Indeed, in the hard turbulence regime,  the typical length scales of convection, such as the plumes various characteristic sizes, are known to decrease continuously as $Ra$ increases, yielding to an increase of the number of degrees of freedom in a given cell \cite{AumaitreEPL2003}.
In physical systems, one expects the relative fluctuations of global variables to decrease as the number $n$ of degrees of freedom contributing to this variable increases \footnote{For example, when the central limit theorem is applicable, the decrease scales like $n^{-1/2}$}, which is not the case of $P^\star$ for $Ra>10^{12}$. To make this argument more concrete, we present below an illustrative model for boundary layer noise in the hard turbulence regime and we show that it leads to a contradiction if we apply it to the high Ra region. 

Our model assumes that $\delta T_{bl}$ fluctuations can be described as resulting from the shot noise of plumes across the boundary layer, where plumes are defined  as independent coherent structures erupting from the plates and responsible for all the heat flux transport. By analogy with the electronic shot noise crossing barriers (for example see \cite{blanter:2000, roche:2005MesoFCS}), shot noise from plumes of heat content $q_0$ is accounted by a white noise source of heat flux with a power spectral density $2 q_0 J$, which is in parallel with the boundary layer resistance $R$. At the low frequency of interest, the inertia $C$ of the plate can be neglected as shown earlier ($R C F_{c} \ll1$). Thus we predict a white noise spectrum for $\delta T_{bl}$ with a power spectral density $P_{bl} = R^2 2 q_0 J$. For consistency, the plateau extent $F_c$ should be to smaller or equal to the inverse of the autocorrelation time of each plume. Since $F_c$ corresponds to 2 turn over times typically, this requirement is not stringent. With this model, it is possible to infer the number $n$ of independent plumes arising from the plate at a given time, which can be seen as the number of degrees of freedom of the boundary layer. We assume that $q_0$ corresponds to a $n^{th}$ of the thermal boundary layer heat content $Q$ with :
\begin{equation}
Q = \rho c_p \frac{\Delta}{4} S \lambda
\end{equation}
\noindent  where $S=\pi h^2 / 16$ and $\lambda \simeq h / 2Nu$ are the surface
and thickness of the boundary layer in the hard turbulence regime. Neglecting the 1.3 prefactor in Eq.\ref{resdyn}, a bit of algebra gives :
\begin{equation}
n=\frac{Pr} {16 P^\star_{bl} Nu^2}
\end{equation}
\noindent where   $P^\star_{bl}=P_{bl} . \nu / (h^2 \Delta^2) $.
For $Ra>2.10^{12}$, the bulk contribution $P^\star$ is reflected so we have $P^\star \sim P^\star_{bl}$ and we note that this value is roughly independent of $Ra$. 
 Then, the above formula indicates that the number of degrees of freedom $n$ of the boundary layer scales as $Nu^{-2}$, that is a decreasing function of $Ra$. This is in contradiction with experience in the hard turbulence regime \cite{AumaitreEPL2003}, which weakens the initial hypothesis of the existence of "hard-turbulence" boundary layers above $Ra \simeq 10^{12}$.

As a final argument, we note that the strong departure from gaussianity for $Ra > 10^{12}$ is not consistent with the continuous increase of the number of degrees of freedom of the boundary layer in the hard turbulence regime. Indeed, such an increase should favor the convegence to a gaussian of the pdf of global quantities such as  $P^\star \sim P^\star_{bl}$, which is not the case.


\section{Conclusions}

A constant heating is applied to the bottom plate of a Rayleigh-B\'enard cell and this plate's temperature fluctuations are analysed. We focus on times scales larger than typically 2 turn over times of the large scale circulation. Below $Ra \simeq 10^{12}$, we give evidence that the plate's fluctuations follow the temperature fluctuations of the bulk of the cell. This implies that the temperature drop across the bottom boundary layer has negligible low frequency fluctuations. Above $Ra \simeq 2 \times 10^{12}$, the bottom plate fluctuation abruptly increase compared to the fluctuations in the bulk. This change is interpreted as an extra noise contribution brought in by the boundary layer. We  discussed and discarded the possibility that this noise was present but simply masked at lower $Ra$. The negative skewness of the plate's temperature -which absolute value quickly increases in the high $Ra$ regime- supports the picture of a boundary layer becoming unstable for $Ra \sim 10^{12}$

The transition of the heat transfer law has already been reported with this cell \cite{chavanne1997}. These authors interpreted this as the triggering of Kraichnan's Ultimate Regime of convection, which is associated with a laminar to turbulent transition of the boundary layer \cite{Kraichnan1962}. The present results are fully consistent with such an interpretation and they provide new evidences supporting it based on dynamical measurement conducted ``on'' the boundary layer.

\acknowledgments 
We thank B. Chabaud, B. H\'ebral, S. Aumaitre, P. Diribarne, Y. Gagne,  and more especially B. Castaing for discussions and inputs. We thank the authors of \cite{chavanne1997} for sharing their raw data. This work was made possible thanks to the R\'egion Rh\^one-Alpes support under contract 301491302.


\end{document}